\documentclass[12pt, twocolumn]{openjournal}
\usepackage{float}
\usepackage{graphicx}
\usepackage{xcolor}
\usepackage{textgreek}
\usepackage[utf8]{inputenc}
\usepackage[english]{babel}
\usepackage{multirow}
\usepackage{hyperref}
\hypersetup{
    colorlinks=true,
    linkcolor=blue,
    filecolor=blue,      
    urlcolor=blue,
    citecolor=blue,
}
\usepackage{color,colortbl}
\usepackage{tensind}
\tensordelimiter{?}
\DeclareGraphicsExtensions{.bmp,.png,.jpg,.pdf}
\usepackage{verbatim}
\usepackage[normalem]{ulem}
\usepackage{orcidlink}
\usepackage{soul}
\usepackage{amsmath}
\graphicspath{ {./figs/} }

\begin{document}
\title{Do We Know How to Model Reionization?}
\author{Nickolay Y. Gnedin\orcidlink{0000-0001-5925-4580}}
\affiliation{Theory Division; 
Fermi National Accelerator Laboratory;
Batavia, IL 60510, USA}
\affiliation{Kavli Institute for Cosmological Physics;
The University of Chicago;
Chicago, IL 60637, USA}
\affiliation{Department of Astronomy \& Astrophysics; 
The University of Chicago; 
Chicago, IL 60637, USA}

\begin{abstract}
I compare the power spectra of the radiation fields from two recent sets of fully coupled simulations that model cosmic reionization: "Cosmic Reionization On Computers" (CROC) and "Thesan". While both simulations have similar power spectra of the radiation sources, the power spectra of the photoionization rate are significantly different at the same values of cosmic time or the same values of the mean neutral hydrogen fraction. However, the power spectra of the photoionization rate can be matched at large scales for the two simulations when the matching snapshots are allowed to vary independently. I.e., on large scales, the clustering of the radiation field in two simulations evolves similarly, but the exact timing of this evolution is different in different simulations and is not parameterized by an easily interpretable physical quantity like the mean neutral fraction or the mean free path. On small scales, large differences are present and remain partially unexplained.

Both CROC and Thesan use the Variable Eddington Tensor approximation for modeling radiative transfer, but adopt different closure relations (optically thin OTVET versus M1). The role of this key difference is tested by using smaller simulations with a new cosmological simulation code that implements both closure relations in a controlled environment (the same hydro, cooling, and gravity solvers and the star formation recipe). In these controlled tests, both the M1 closure and the OTVET ansatz follow the expected behavior from a simple analytical approximation, demonstrating that the differences in the 2-point function of the radiation field induced by the choice of the Eddington tensor are not dominant.
\end{abstract}


\maketitle

\section{Introduction}
\label{sec:intro}

One of the primary challenges in modeling the real universe is that it is often very hard to validate numerical codes - different codes may all pass simple tests perfectly, and yet diverge in realistic simulations. Hence, "comparison projects" have been used extensively to validate codes that use very different numerical approaches by modeling the same physical system, beginning with the pioneering "Santa Barbara Cluster Comparison" project \citep{Frenk1999} and the first N-body code comparisons \citep{Knebe2000}. 

For modeling cosmic reionization, an accurate radiative transfer solver is especially important. "Iliev tests" \citep{iliev1,iliev2} serve as the de facto standard for cosmological radiative transfer codes, but they test radiative transfer solvers and their coupling to hydrodynamics in restricted idealized settings. So far, no inter-comparison of different simulations of reionization, similar to existing comparisons of other cosmological codes like AGORA \citep{agora}, has been done. 

\begin{figure}[t]
\centering
\includegraphics[width=\columnwidth]{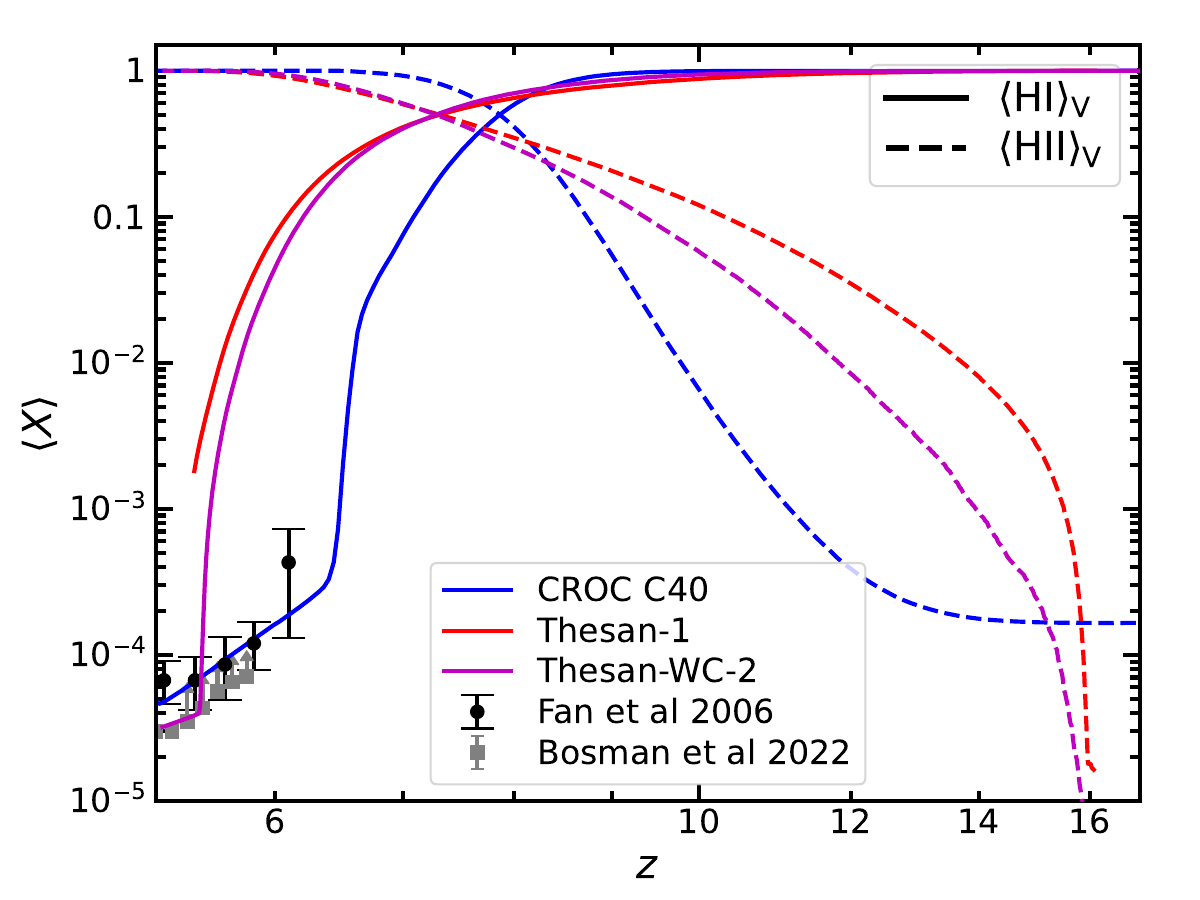}\hfill
\caption{Volume weighted neutral (solid lines) and ionized (dashed lines) fraction of hydrogen as functions of redshift for CROC (blue) and Thesan (red) runs used in this work. The data points from \citet{Fan2006} and \citet{Bosman2022}  were used as the calibration data for CROC and Thesan simulations, respectively. I also show in magenta Thesan-WC-2, one of Thesan's calibration runs.}
\label{fig:xhz}
\end{figure}

\begin{figure*}[t]
\centering
\includegraphics[width=\textwidth]{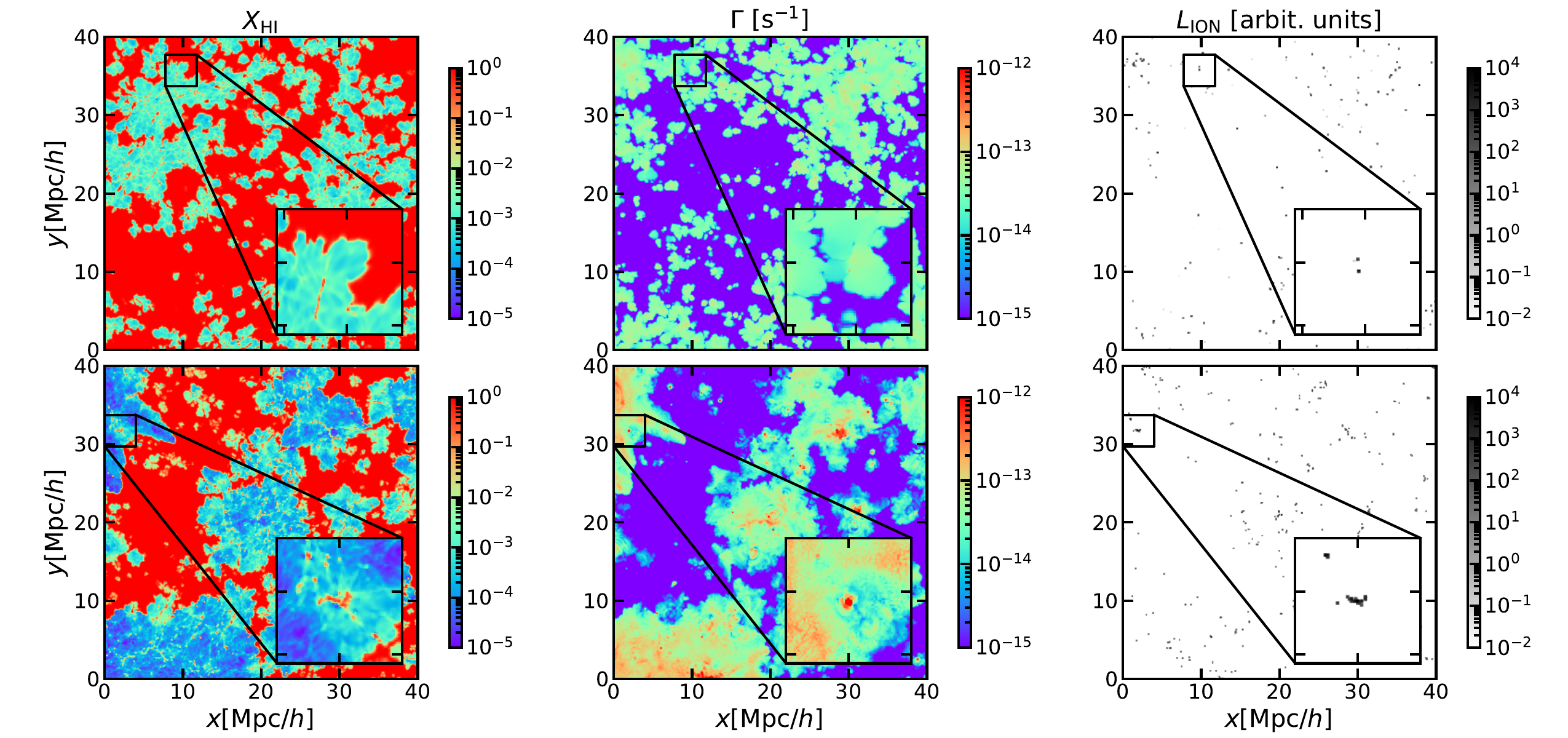}\hfill
\caption{Slices through CROC "C40A" (top) and Thesan-1 (bottom) simulations at times when each simulation is approximately at mid-point of reionization ($\langle X_{\rm HI}\rangle_V \approx 0.5$). Three panels show the neutral fraction, the photoionization rate, and the radiation source (in arbitrary units). The inserts are centered on the strongest peak in the photoionization rate within these randomly chosen slices. The Thesan box is cropped to match the size of a smaller CROC box. \label{fig:slice}}
\end{figure*}

Two simulation sets that serve as representative examples of the current "state-of-the-art" are publicly available:  "Cosmic Reionization On Computers" \citep[CROC,][]{Gnedin2014,GnedinKaurov2014,Gnedin2022} and "Thesan" \citep{Kannan2022,Garaldi2022,Garaldi2024}. The two sets of simulations are sufficiently similar so that a reasonable agreement can be expected between them. They model similar $\sim 100$ cMpc volumes with similar spatial resolutions of 100-300 proper parsecs, have almost identical mass resolution ($2000^3$), and model a similar range of relevant physical effects: hydrodynamics, non-equilibrium cooling and ionization, radiative transfer, and star formation and stellar feedback. In modeling the latter, Thesan does a noticeably better job than CROC, as CROC only includes star formation in molecular gas and has been shown to underpredict star formation rates and UV luminosities of most massive galaxies \citep{Zhu2020}, while Thesan uses BPASS for stellar spectra, and includes binary stars, which emit ionizing photons up to 3 times longer than single stars.

Despite their similarities, the two simulation sets have sufficiently different reionization histories, shown in Figure \ref{fig:xhz}. Since both simulation projects match the observed evolution of the dust-corrected galaxy luminosity function at $z<10$ reasonably well, the difference between them is not due to a difference in how sources of ionizing radiation are modeled (this is also illustrated quantitatively below in Figure \ref{fig:gam}), but rather in how ionizing radiation propagates through the simulation volume and how (and where) it is absorbed. Both simulation projects use the Variable Eddington Tensor approximation for modeling radiative transfer on a grid with a moments method, adopt a "reduced speed of light" approximation \citep{Gnedin2001} for evolving the radiation field forward in time, and feature the "effective escape fraction" to account for photon absorptions below the spatial resolution of the simulations. The main difference between the two projects is in different ansatzes for the Variable Eddington tensor - this may serve as one reason for the difference between them. Another major difference is in the actual equations they are solving. Both these differences and several other, less obvious numerical effects are explored in this paper.

\section{Simulations}
\label{sec:sims}

As has already been mentioned, the two simulation sets ("CROC" and "Thesan" hereafter) have sufficiently different reionization histories. Hence, the comparison between them is not trivial; the corresponding snapshots between the two simulations must be "matched" somehow. One possible way to match them is to compare snapshots at the same value of the volume-weighted mean neutral fraction, $\langle X_{\rm HI}\rangle_V$ (the mass-weighted fraction during and after the overlap of ionized bubbles is dominated by DLAs and is not representative of the ionization state of the IGM). Figure \ref{fig:slice} shows a visualization of the two simulations at around the "mid-point" of reionization when $\langle X_{\rm HI}\rangle_V\approx0.5$. Thus matched, the two simulations are not drastically different: the distribution of sources is similar, ionized bubble morphology is also similar (although hydrogen is more ionized inside the bubbles in Thesan than in CROC, and CROC has more small bubbles, again reflecting its deficiency in modeling the largest galaxies), and the largest difference seems to be in the distribution of the photoionization rate inside the bubbles. Exploring this difference is the key goal of this paper.

Publicly available Thesan data include uniform grids with a number of relevant physical fields, with a small twist: the ionizing luminosity is binned in redshift, not real space. Since the clustering in redshift space is suppressed on small scales compared to real space, that particular choice of the Thesan team complicates the comparison between the radiation source and the actual radiation field. Hence, hereafter, both the ionizing luminosity in redshift space and the stellar density in real space are used as proxies for the Thesan radiation source in real space.

\begin{figure*}[t]
\centering
\includegraphics[width=\columnwidth]{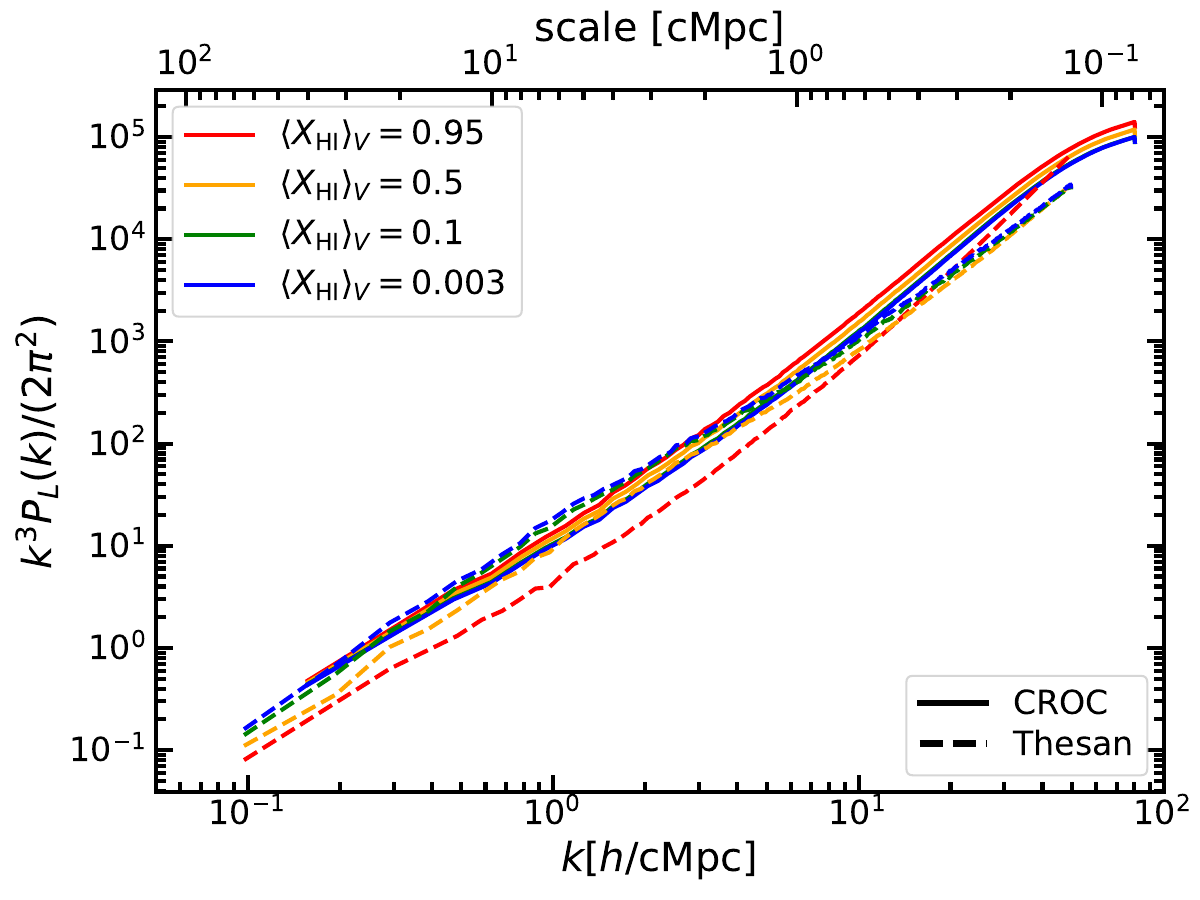}\hfill
\includegraphics[width=\columnwidth]{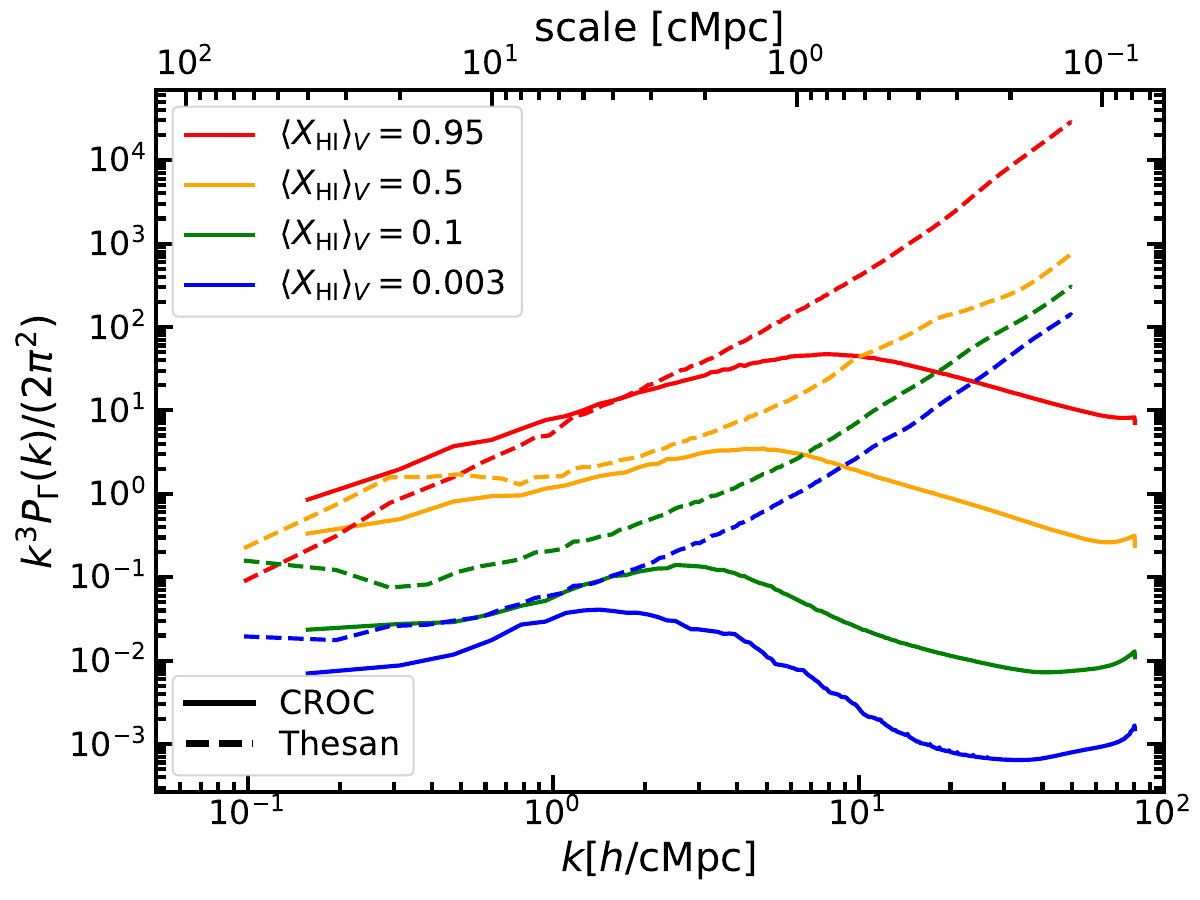}%
\caption{Power spectra of the radiation source density (left) and the photoionization rate (right) in CROC (solid) and Thesan (dashed) simulations at four matched times when each simulation has a given value of the volume-weighted neutral hydrogen fraction (given in the legend).}
\label{fig:gam}
\end{figure*}

Existing CROC data do not include uniform grids with the radiation source or the radiation field. These can be produced from the full simulation snapshot data, but doing so requires loading the full simulation snapshot into computer memory, and there are no readily available computational resources for operating on the largest, $80h^{-1}$ cMpc CROC boxes. Hence, for CROC smaller, $40h^{-1}$ cMpc boxes are used. CROC produced six independent random realizations of this box that differ only by the values of random amplitudes in the initial conditions, including the random amplitude of the fundamental wave, the so-called "DC mode" \citep{Sirko2005}. These six realizations are labeled "C40A"-"C40F", and in order to save the limited computational resources, only three of them ("C40A", "C40C", and "C40F") are used in this work. The average ionization history of these three is virtually indistinguishable from the average reionization history of the full six-simulation set, so they serve as a good representation of the entire set. In addition, the three simulations sample the spatial volume of $3\times (40/h)^3 = (58/h)^3$ cMpc$^3$, only moderately lower than the volume of the Thesan-1 simulation of $(65/h)^3$ cMpc$^3$. However, even taken together, the three $40h^{-1}$ cMpc boxes do not sample the scales larger than their fundamental frequency, so Thesan reaches $\approx 60$\% larger scales than CROC.

Another potentially important difference between CROC and Thesan is that the former includes a second, "background" component of the radiation field designed to account for radiation exiting and entering the simulation box and for radiation emitted by quasars, which are too rare to be captured in CROC and Thesan simulation volumes. In order to avoid complications arising from this difference, in this comparison, I exclude this background component so that the comparison between CROC and Thesan is as direct as possible.

\section{Results}
\label{sec:res}

There are several major differences in physical modeling between CROC and Thesan: models of star formation and stellar feedback, cooling and heating functions, approaches for accounting for the AGN contribution to ionizing sources, inclusion of explicit dust modeling in Thesan, etc. Therefore, perhaps surprisingly, the power spectra of the radiation source density, shown in the left panel of Figure \ \ref{fig:gam}, are very similar. The power spectra of the hydrogen photoionization rate, shown in the right panel of the same figure, are, however, very different. Not only do the CROC simulations exhibit a suppression of power on small scales, but on large scales, while the slopes of the power spectra between the two simulations approximately match, the amplitudes do not. 

In the Newtonian limit (the speed of light $c \rightarrow \infty$) and ignoring the Hubble expansion (a valid assumption for CROC and Thesan boxes whose sizes are much smaller than the Hubble radius) the relationship between the comoving radiation source density $L(\vec{x})$ and the radiation energy density $E(\vec{x})$ at a particular frequency can be expressed in a closed form as
\begin{equation}
    E(\vec{x}) = \frac{a}{4\pi} \int d^3 x^\prime \frac{L(\vec{x}^\prime)}{(\vec{x}-\vec{x}^\prime)^2} e^{-\tau(\vec{x},\vec{x}^\prime)},
    \label{eq:rt}
\end{equation}
where $\tau(\vec{x},\vec{x}^\prime)$ is the optical depth between comoving locations $\vec{x}$ and $\vec{x}^\prime$. Hence, in the most general case, the clustering of the radiation field is determined by both the clustering of sources and by clustering of the opacity. 

Generally, equation (\ref{eq:rt}) is not a convolution. In a special case of constant opacity 
\[
    \tau(\vec{x},\vec{x}^\prime) = \frac{|\vec{x}-\vec{x}^\prime|}{\lambda},
\]
where $\lambda$ is the (spatially constant) mean free path. In that special case, the power spectrum of the radiation field $P_E(k)$ and the radiation source density power spectral $P_L(k)$ are directly proportional to each other,
\[
    P_E(k) = P_L(k) T^2_k,
\]
where
\begin{equation}
    T_k = \alpha\frac{\arctan(k\lambda)}{k\lambda}
    \label{eq:tk}
\end{equation}
and $\alpha=\lambda\langle L\rangle/\langle E\rangle$ is around unity during reionization and is somewhat lower after the overlap.

Opacity is, of course, not constant generally. It varies significantly in the ISM and CGM of ionizing sources on sub-kpc scales, but these variations can be accounted for as variations in the source escape fractions and occur on scales much smaller than those shown in Figure \ \ref{fig:gam}. In the IGM on average $\tau(\vec{x},\vec{x}^\prime) \ll 1$ on scales much smaller than the mean free path, so after the overlap, when the mean free path becomes comparable to the box sizes of CROC and Thesan,  one may plausibly still expect the asymptotic behavior $T_k\propto k^{-1}$.

\begin{figure}[t]
\centering
\includegraphics[width=\columnwidth]{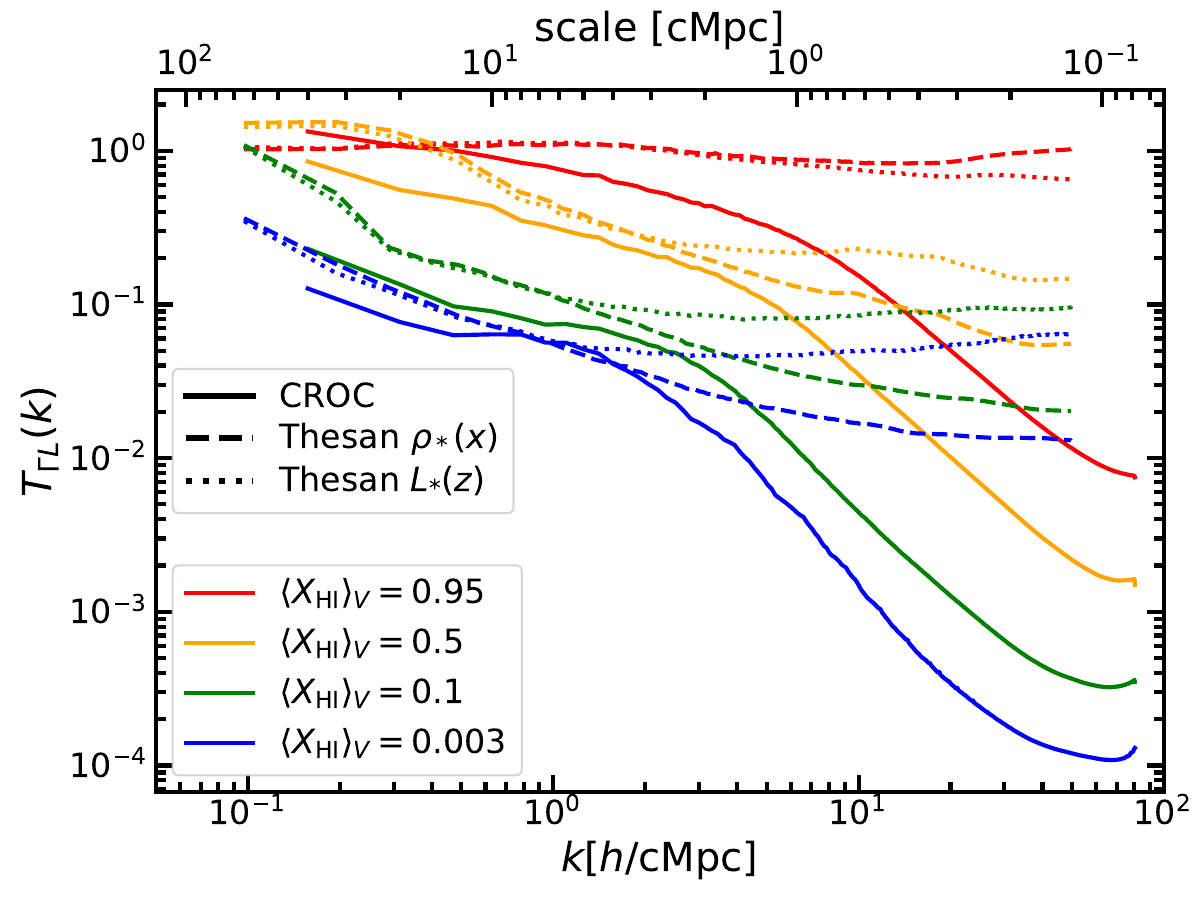}\hfill
\caption{The photoionization rate "transfer function" $T_{\Gamma L}$ in CROC (solid) and Thesan (dashed and dotted) simulations at four matched times when each simulation has a given value of the volume-weighted neutral hydrogen fraction (given in the legend). For Thesan, both the mass density of stars in real space (dashed lines) and the ionizing luminosity in redshift space (dotted) are used as proxies for the radiation source.}
\label{fig:tk}
\end{figure}

This relation serves as a motivation to explore the "transfer function" between the photoionization rate and the radiation source density, $T_{\Gamma L}(k) \equiv (P_\Gamma(k)/P_L(k))^{1/2}$. The transfer functions for the snapshots from Figure \ref{fig:gam} are shown in Figure \ref{fig:tk}. The differences between CROC and Thesan on small scales are equally apparent in this figure, and these differences are much larger than the differences between the two choices of the radiation source in the Thesan simulation. In addition, neither of the two simulations exhibits the small-scale behavior $T_{\Gamma L}(k)\propto k^{-1}$.

The differences between the two choices of the radiation source in Thesan simulation only manifest themselves on small scales, as can be expected from the difference between the real and redshift space power spectra. Hence, hereafter, the stellar density (in real space) is used as a proxy for the radiation source density in Thesan.

One can notice, though, that matching by the value of the mean neutral fraction is somewhat arbitrary - if reionization proceeds differently in two simulations, there is no reason to expect that the same values of the mean neutral fraction correspond to agreements between other statistical descriptions of the reionization process, such as the distribution of bubble sizes or, in our case, the clustering of the radiation field. For example, the transfer function in CROC at $\langle X_{\rm HI}\rangle_V \approx 0.1$ matches the Thesan transfer function on large scales at $\langle X_{\rm HI}\rangle_V \approx 0.003$. One can wonder, therefore, whether a better match between the two simulations can be achieved if the matching is done not by the value of the volume-weighted mean neutral fraction but by the transfer function itself - i.e. if the Thesan snapshot is selected to match the transfer function from the CROC snapshot as best as possible.

\begin{figure}[t]
\centering
\includegraphics[width=\columnwidth]{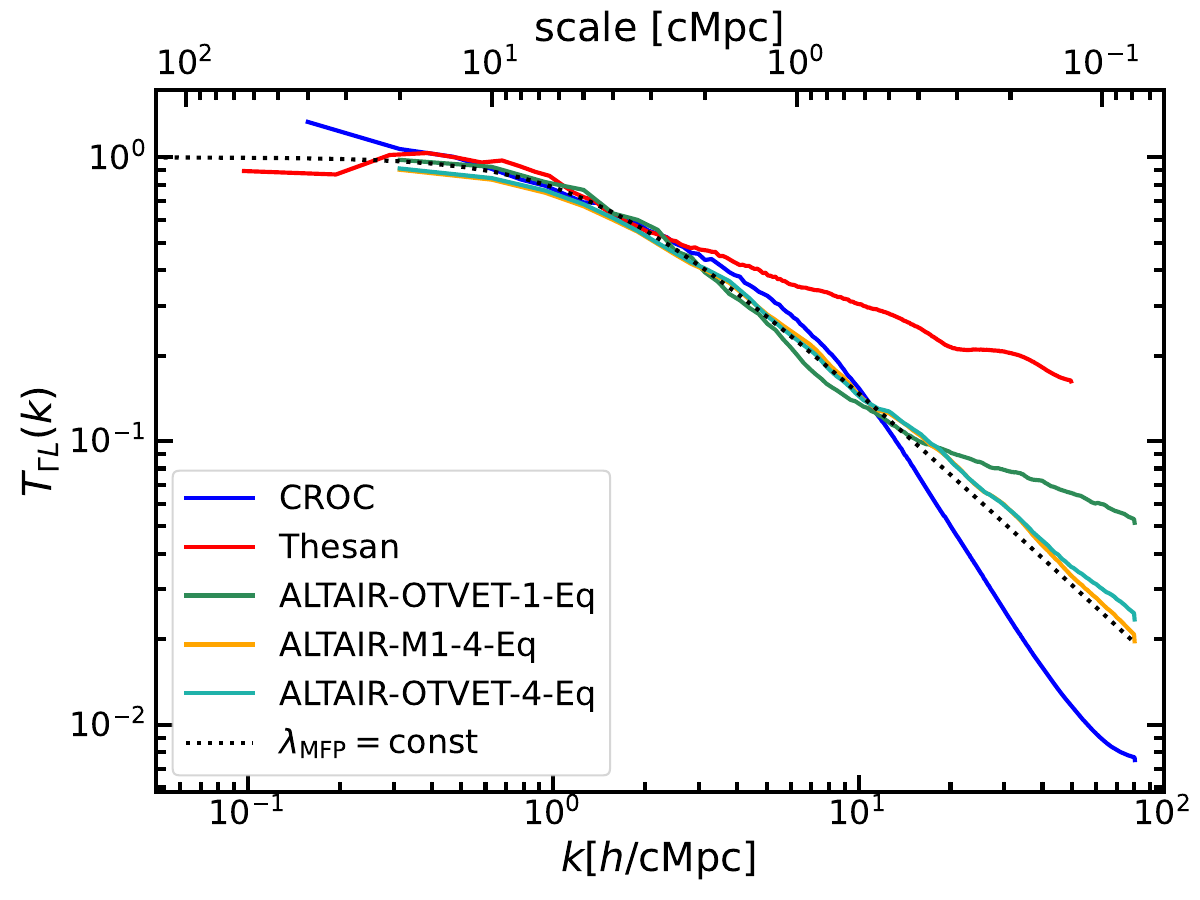}\hfill
\caption{The photoionization rate "transfer function" $T_{\Gamma L}$ in CROC, Thesan, and several ALTAIR simulations at best matching times for a representative example (as described in the text). In CROC at this moment $\langle X_{\rm HI}\rangle_V=0.95$ and in Thesan $\langle X_{\rm HI}\rangle_V=0.7$. The black dotted line shows an analytical calculation for constant mean free path $\lambda$, $T_{\Gamma L} = \arctan(k\lambda)/(k\lambda)$ for the best matching $\lambda$. }
\label{fig:alt4}
\end{figure}

Figure \ref{fig:alt4} shows thus matched transfer functions for CROC and Thesan as blue and red lines, respectively.  For the chosen CROC snapshot with $\langle X_{\rm HI}\rangle_V=0.95$, this corresponds to the Thesan snapshot with $\langle X_{\rm HI}\rangle_V=0.7$, and matches for other snapshots are shown in the Appendix \ref{app:a}. The figure also shows the "constant opacity" transfer function $T_{\Gamma L} = \arctan(k\lambda)/(k\lambda)$ with the value of $\lambda$ adjusted manually to approximately fit the simulation results on large scales.

It is difficult to understand the differences between the two simulation sets further by only analyzing the simulation results. There are two obvious differences, and there may be other, more obscure differences as well. The first obvious difference is that CROC uses the Optically Thin Variable Eddington Tensor (OTVET) closure for the radiation moment hierarchy \citep{Gnedin2001} while Thesan uses a so-called "M1 closure" \citep{Pomraning1969} as described in \citet{Kannan2019}. Hence, the two approaches adopt different approximations for the radiation Eddington Tensor. Additionally, the OTVET approximation as implemented in CROC folds 4 equations for the radiation energy density and flux into one. 

Namely, equations for the first two moments of the radiation field (in a comoving reference frame and with cosmological terms neglected within a simulation volume) are
\begin{subequations}%
\label{eq:rtmoms}%
\begin{align}
    \frac{a}{c}\frac{\partial E}{\partial t} + \frac{\partial F^j}{\partial x^j} & = 
    -\kappa E + L, \\
    \frac{a}{c}\frac{\partial F^i}{\partial t} + \frac{\partial P^{ij}}{\partial x^j} & = 
    -\kappa F^i + Q^i, 
\end{align}
\end{subequations}
where $F^i$ is the radiation flux, $P^{ij} \equiv E h^{ij}$ is the radiation pressure tensor, $h^{ij}$ is the Eddington tensor, and $Q^i$ is the flux source, which vanishes for isotropic sources. 

In the Newtonian limit (i.e.\ in the limit $c\rightarrow\infty$), the time derivative in the flux equations can be dropped \citep{Gnedin2001},
\[
    F_i = -\frac{1}{\kappa} \frac{\partial E h^{ij}}{\partial x^j},
\]
and hence
\begin{equation}
    \frac{a}{c}\frac{\partial E}{\partial t} = \frac{\partial}{\partial x^i} \frac{1}{\kappa+\epsilon} \frac{\partial }{\partial x^j}E h^{ij} - \kappa E + L,
    \label{eq:diff}
\end{equation}
where a small number $\epsilon$ is added to avoid occasional numerical overflow. OTVET implementation in the ART code \citep{art1,art2,art3}, used to run CROC simulations, relies on this equation to solve for $E$ given the Eddington tensor $h^{ij}$.

In order to explore the effects of these approximations, I use a new cosmological simulation code that implements all possible combinations of these approximations. The code's working name is ALTAIR (which stands for "Adaptive, Locally Timestepping, Asynchronous Implementation of Refinement"). It is still currently in development and will be described in detail elsewhere. Here, it is only mentioned that the code is an implementation of the Adaptive Mesh Refinement methodology, and it implements all the physics that CROC simulations include. Namely, gravity is solved for with an FMM gravity solver from \citet{Gnedin2021} that supports periodic boundary conditions natively, hydro is solved with a highly robust hydro solver from the RAMSES code \citep{Teyssier2002}, cooling, chemistry, and star formation/feedback are modeled with the same approach as used in CROC \citep{Gnedin2014}. ALTAIR features local time-stepping and full enforcement of the Courant-Friedrichs-Lewy condition as described in \citet{Gnedin2018}.

Limitations on the available computational resources mean that only small, lower-resolution boxes are run for this work. Three ALTAIR simulations that are used in this work have $20 h^{-1}$ cMpc boxes and $256^3$ dark matter particles, and a spatial resolution of only 7 comoving kpc (3 times worse than Thesan and 10 times worse than CROC). The purpose of these simulations is not to produce accurate models of reionization but merely to compare different radiative transfer methods in a controlled environment - within the same simulation with identical gravity and hydro solvers and cooling and star formation models.

These three additional simulations are shown in Figure \ref{fig:alt4} with greenish-orange lines. "ALTAIR-OTVET-1-Eq" is meant to be a direct analog of CROC: using OTVET approximation and a single equation (\ref{eq:diff}) for the radiation intensity. "ALTAIR-M1-4-Eq" is meant to be a direct analog of Thesan: using the M1 closure and four equations (\ref{eq:rtmoms}). Finally, "ALTAIR-OTVET-4-Eq" is an intermediate case: using four equations (\ref{eq:rtmoms}) and the identical numerical solver to "ALTAIR-M1-4-Eq", but using the optically thin Eddington tensor instead of the M1 closure (the fourth combination, the M1 closure and one equation, is not possible). All three forms pass several Iliev tests \citep{iliev1,iliev2}, as documented in Appendix \ref{app:c}.

Comparing all five simulations from Figure \ref{fig:alt4} is both encouraging and disappointing. While there is a reasonable agreement among all five simulations on linear and quasi-linear scales ($\gtrsim 2h^{-1}$ cMpc), the discrepancies on small scales are large and not easily interpretable. First, Thesan exhibits much larger power on small scales, and the origin of that power is not clear. One possible reason is a more stochastic star formation rate in Thesan galaxies that may result in a weaker correlation between the instantaneous values of the radiation field at one location and the radiation source at another. Namely, Equation (\ref{eq:rt}) in the next order in $1/c$ becomes

\begin{align}
    E(\vec{x},t) & = \frac{a}{4\pi} \int d^3 x^\prime \frac{L(\vec{x}^\prime,t-|\vec{x}-\vec{x}^\prime)|/c)}{(\vec{x}-\vec{x}^\prime)^2} e^{-\tau(\vec{x},\vec{x}^\prime)} \nonumber\\
    & \approx \frac{a}{4\pi} \int d^3 x^\prime e^{-\tau} \left[\frac{L(\vec{x}^\prime,t)}{(\vec{x}-\vec{x}^\prime)^2}  - \frac{\dot{L}(\vec{x}^\prime,t)}{c |\vec{x}-\vec{x}^\prime|}\right]. \nonumber
\end{align}
For a large mean free path, on small scales ($\tau \ll 1$) the last expression behaves as $L_k/k - \dot{L}_k/k^2$, so the fluctuations in the star formation rate can create a plateau in $T_{\Gamma L}(k)$ only if the star formation rate Fourier transform scales as $k^2 L_k$, which is not the case - both the power spectra of the stellar density and the star formation rate scale on small, nonlinear scales in Thesan as expected, $L_k \propto \dot{L}_k \propto k^{-1/2}$. Another possible guess for the plateau in Thesan $T_{\Gamma L}(k)$ is numerical noise introduced by rebinning from a moving irregular mesh onto a uniform grid, but the spatial scale at which the plateau develops is too large for that.

The transfer function $T_{\Gamma L}(k)$ in CROC, on the other hand, exhibits a suppression at scales below about 1 cMpc. This suppression is due to the highly approximate nature of the optically thin Eddington tensor calculation in CROC. ALTAIR simulations that use an accurate calculation of the optically thin Eddington tensor do not show such a suppression for the same numerical scheme.

The "ALTAIR-OTVET-1-Eq" simulation also exhibits extra small-scale power. This power is due to the $\epsilon$ factor added in Equation (\ref{eq:diff}) - if $\epsilon$ is small ($10^{-3}/\Delta x$ in CROC and $10^{-6}/\Delta x$ in the ALTAIR simulation shown where $\Delta x$ is the local cell size in the simulation), then in regions of low opacity it can amplify numerical noise. The same simulation with $\epsilon=0.01/\Delta x$ does not exhibit the excess of small-scale power, but then it introduces too much artificial absorption that leads to photon non-conservation.

One may wonder if a good agreement between ALTAIR simulations and an analytical model (\ref{eq:tk}) is due to the lower mass resolution of ALTAIR simulations. However, an ALTAIR simulation with the same mass resolution as CROC and Thesan ($256^3$ in a $10/h$ cMpc box) also agrees with the analytical model (\ref{eq:tk}) on small scales, so the agreement between ALTAIR and Equation (\ref{eq:tk}) is not an artifact of low mass resolution.

\begin{figure}[t]
\centering
\includegraphics[width=\columnwidth]{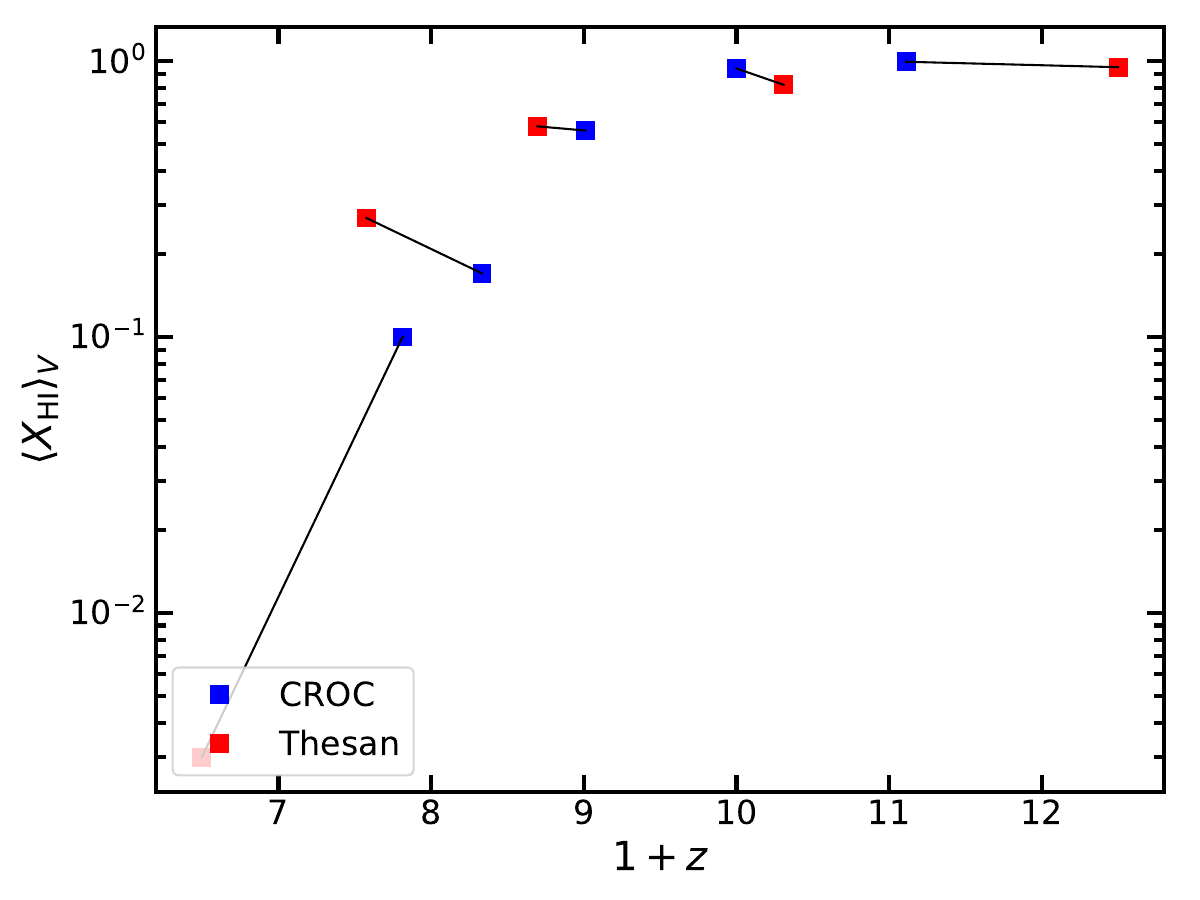}\hfill
\caption{Values for the volume-weighted neutral fraction and redshift for the approximate matches between CROC and Thesan. }
\label{fig:match}
\end{figure}

It would be informative if one could identify how the snapshots in the two simulations can be matched. Figure \ref{fig:match} shows the volume-weighted hydrogen neutral fraction and redshift for the $T_{\Gamma L}(k)$-matched snapshots for CROC and Thesan. Unfortunately, the points from the two simulations do not line up, so there is no obvious way to match them a priori - i.e., matching snapshots by $T_{\Gamma L}(k)$ results in a mismatch in the corresponding values of the neutral fraction or the comoving or physical mean free path ($\approx a^2/\langle X_{\rm HI}\rangle_V$ and $\approx a^3/\langle X_{\rm HI}\rangle_V$ respectively) and vice versa - matching by the neutral fraction of the mean free path does not results in matching $T_{\Gamma L}(k)$.

\section{Conclusions}

In summary, the level of agreement in the spatial clustering of the radiation field between CROC and Thesan, as well as ALTAIR simulations that are designed to mimic them, is not fully satisfactory. On large scales, the clustering of the radiation field in two simulations evolves similarly, but the exact timing of this evolution is different in different simulations and is not parameterized by an easily interpretable physical quantity like the mean neutral fraction or the mean free path. 

Curiously, a high level of matching can also be achieved for other physical properties. While it is somewhat tangential to the focus of this paper, in Appendix \ref{app:b} I show matching between the distributions of effective opacities \citep[mean opacity along skewers of $50 h^{-1}$ cMpc along different lines of sight,][]{Becker2015}. By adjusting the timing in the two simulations, these distributions can be made to match the observational data to within the sample variance.

On small scales, the differences are even larger, with neither CROC nor Thesan following the plausibly expected $T_{\Gamma L} \propto 1/k$ behavior. One good news, however, is that in controlled tests with the ALTAIR code, both the M1 closure and the OTVET ansatz follow the expected behavior, hence demonstrating that the differences in the 2-point function of the radiation field induced by the choice of the Eddington tensor are not dominant.

Hence, the answer to the paper title is "not sure".

\section*{Acknowledgments}

This work was motivated by extensive discussions with Hanjue Zhu, who humbly declined co-authorship in this paper despite her significant role in initiating this work. The paper has also benefited greatly from the discussions at the "Friday Owls" group meeting at the University of Chicago, including Andrey Kravtsov and Harley Katz. I thank Enrico Garaldi for valuable comments and corrections to an earlier version of this manuscript and the Thesan team for making their data publicly available. I am also grateful to the anonymous referee for constructive criticism of the original manuscript. This work was supported by Fermi Forward Discovery Group, LLC under Contract No.\ 9243024CSC000002 with the U.S. Department of Energy, Office of Science, Office of High Energy Physics. This work used resources of the Argonne Leadership Computing Facility, which is a DOE Office of Science User Facility supported under Contract DE-AC02-06CH11357. An award of computer time was provided by the Innovative and Novel Computational Impact on Theory and Experiment (INCITE) program. This research is also part of the Blue Waters sustained-petascale computing project, which is supported by the National Science Foundation (awards OCI-0725070 and ACI-1238993) and the state of Illinois. Blue Waters is a joint effort of the University of Illinois at Urbana-Champaign and its National Center for Supercomputing Applications. I also acknowledge the support from the grant NSF PHY-2309135 to the Kavli Institute for Theoretical Physics (KITP) and the support from the University of Chicago’s Research Computing Center.

\bibliographystyle{mnras}
\bibliography{main}

\vfill
\clearpage

\begin{appendix}

\section{CROC-Thesan Matching at Other Snapshots}
\label{app:a}

\begin{figure}[H]
\centering
\includegraphics[width=0.33\columnwidth]{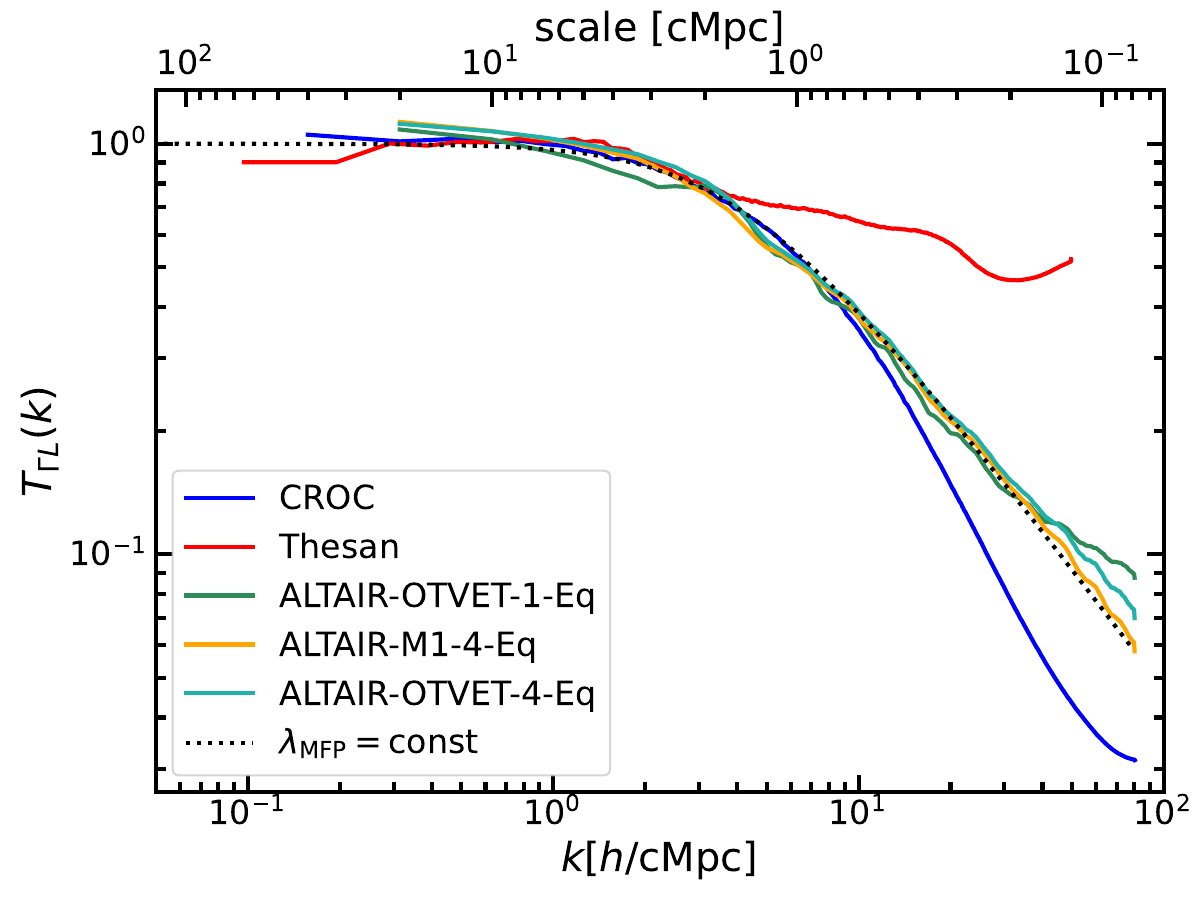}%
\includegraphics[width=0.33\columnwidth]{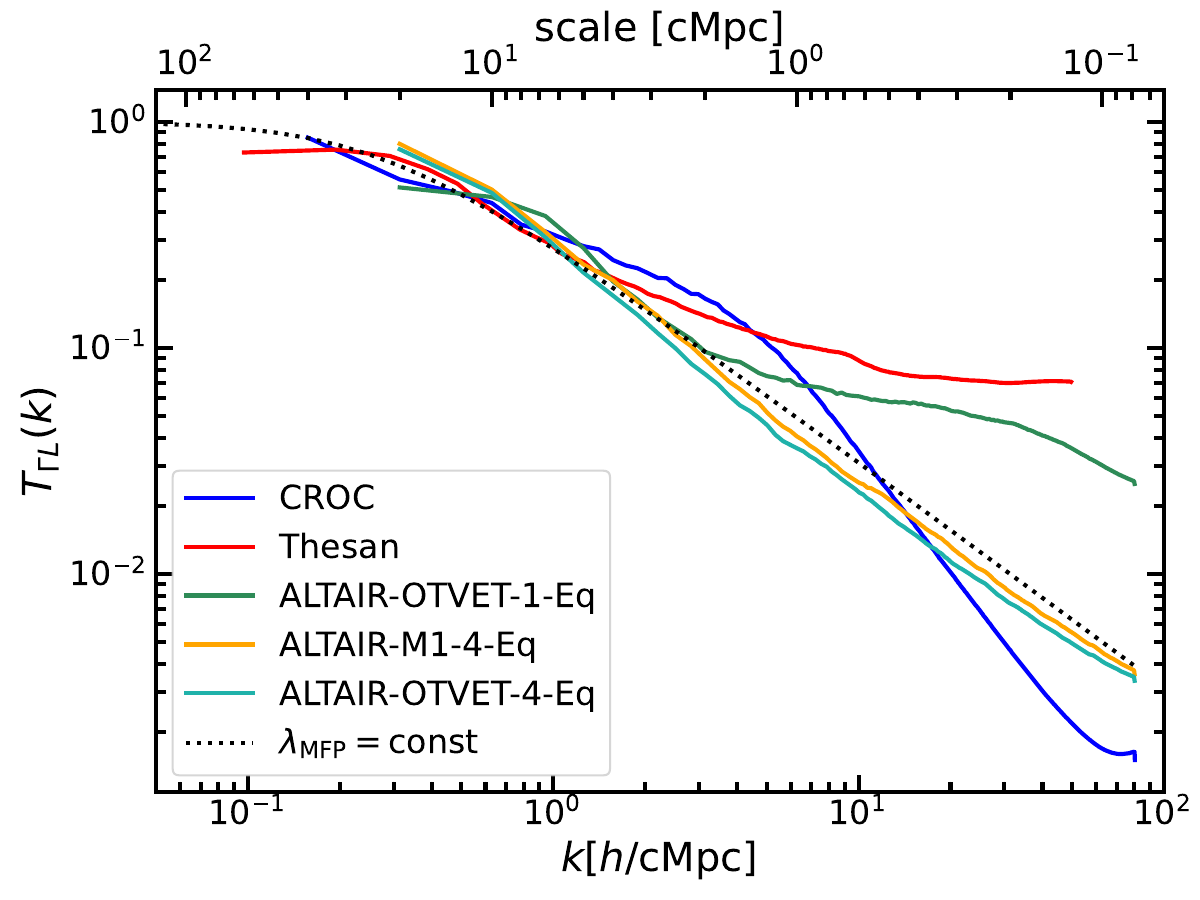}%
\includegraphics[width=0.33\columnwidth]{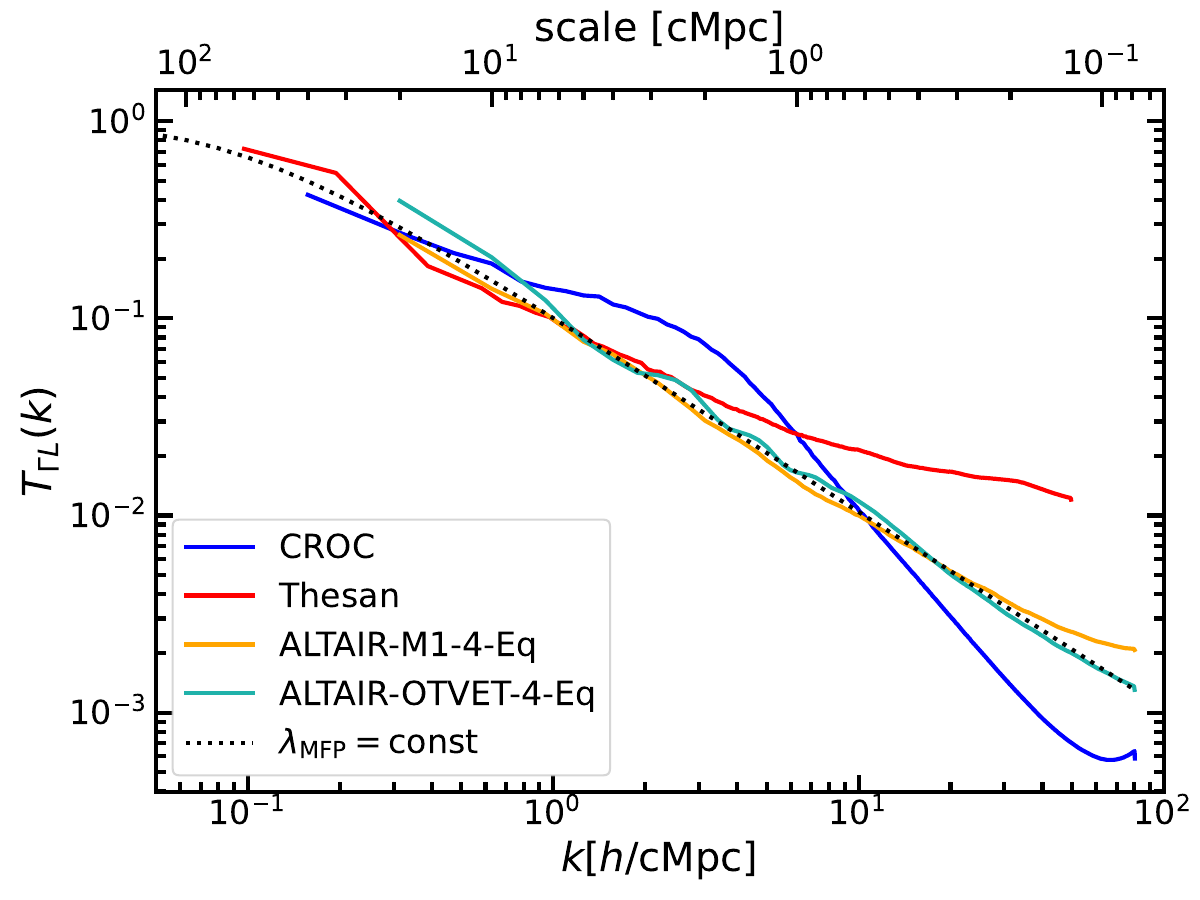}\hfill
\caption{The same as Figure \ref{fig:alt4} at three other matching times: CROC/Thesan $\langle X_{\rm HI}\rangle_V=0.99/0.93$, $0.56/0.36$, and $0.17/0.18$.}
\label{fig:alt4x}
\end{figure}

Figure \ref{fig:alt4x} shows matching between CROC, Thesan, and ALTAIR simulations, similar to Figure \ref{fig:alt4} but at three other moments during reionization.

\section{CROC-Thesan Matching for the Distribution of Effective Opacities}
\label{app:b}

\begin{figure}[H]
\centering
\includegraphics[width=0.25\columnwidth]{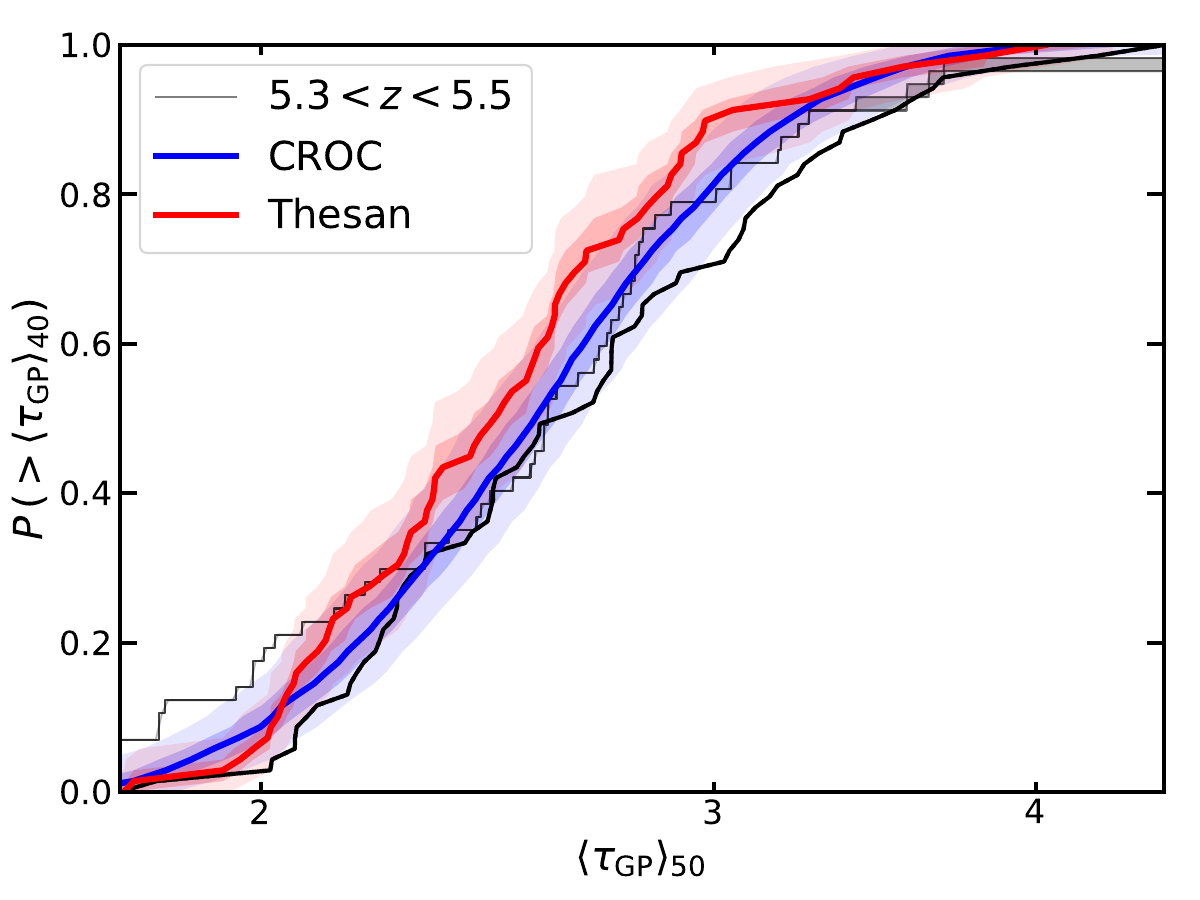}%
\includegraphics[width=0.25\columnwidth]{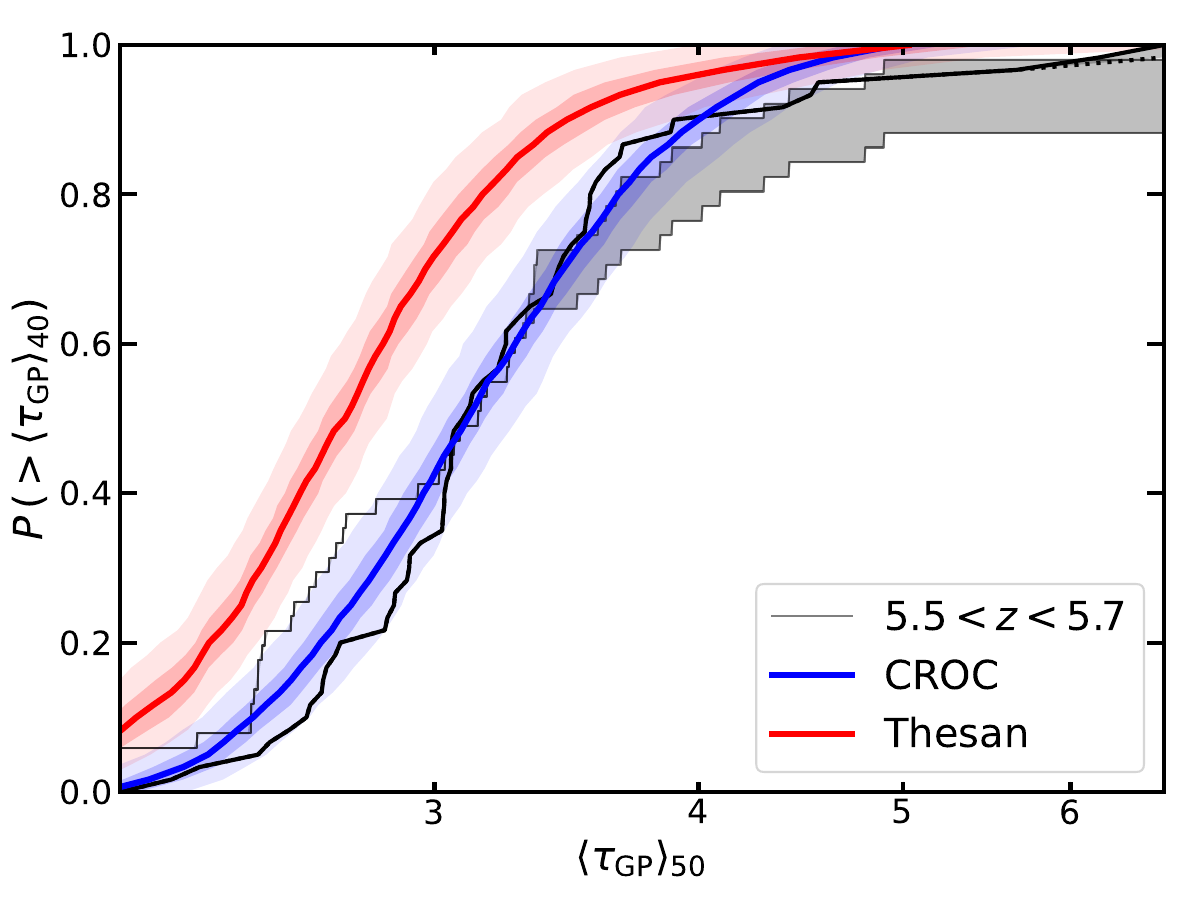}%
\includegraphics[width=0.25\columnwidth]{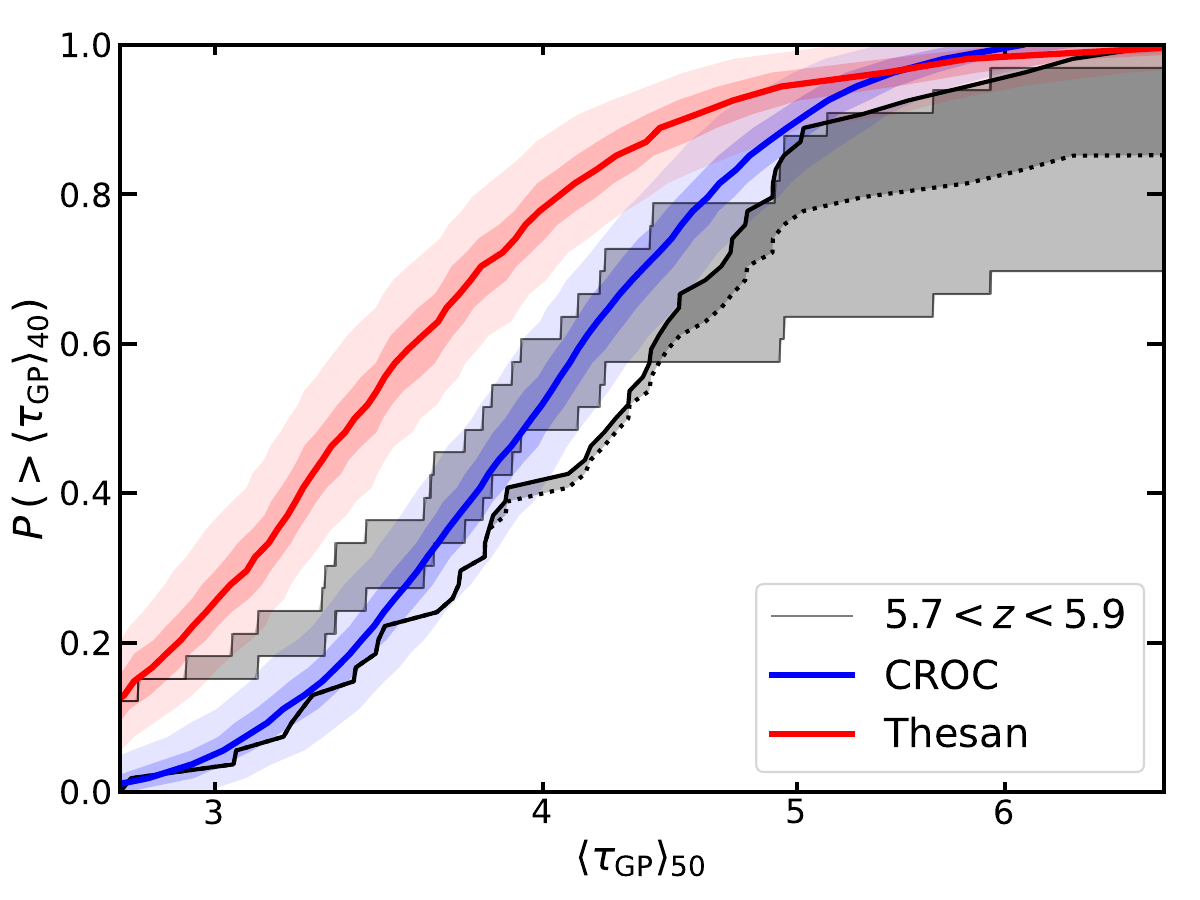}%
\includegraphics[width=0.25\columnwidth]{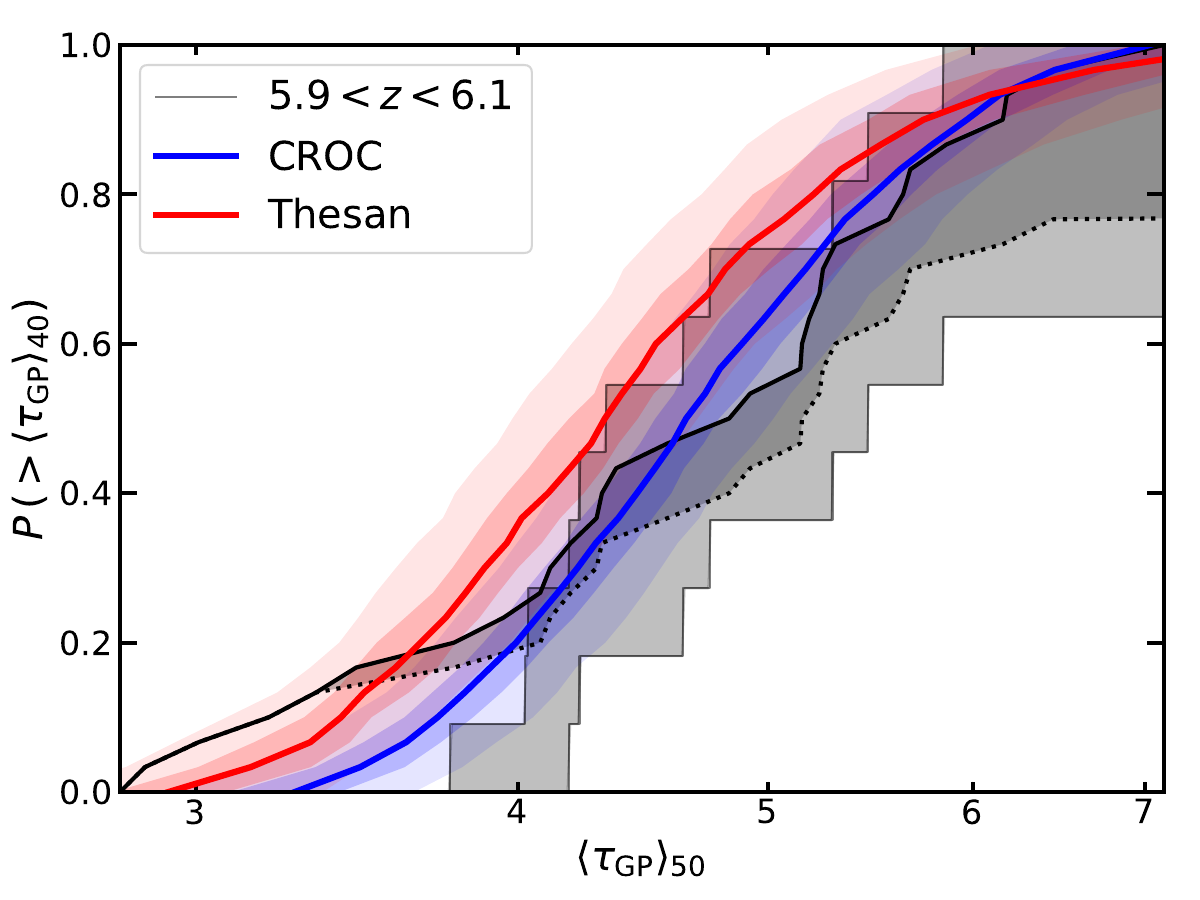}\hfill
\includegraphics[width=0.25\columnwidth]{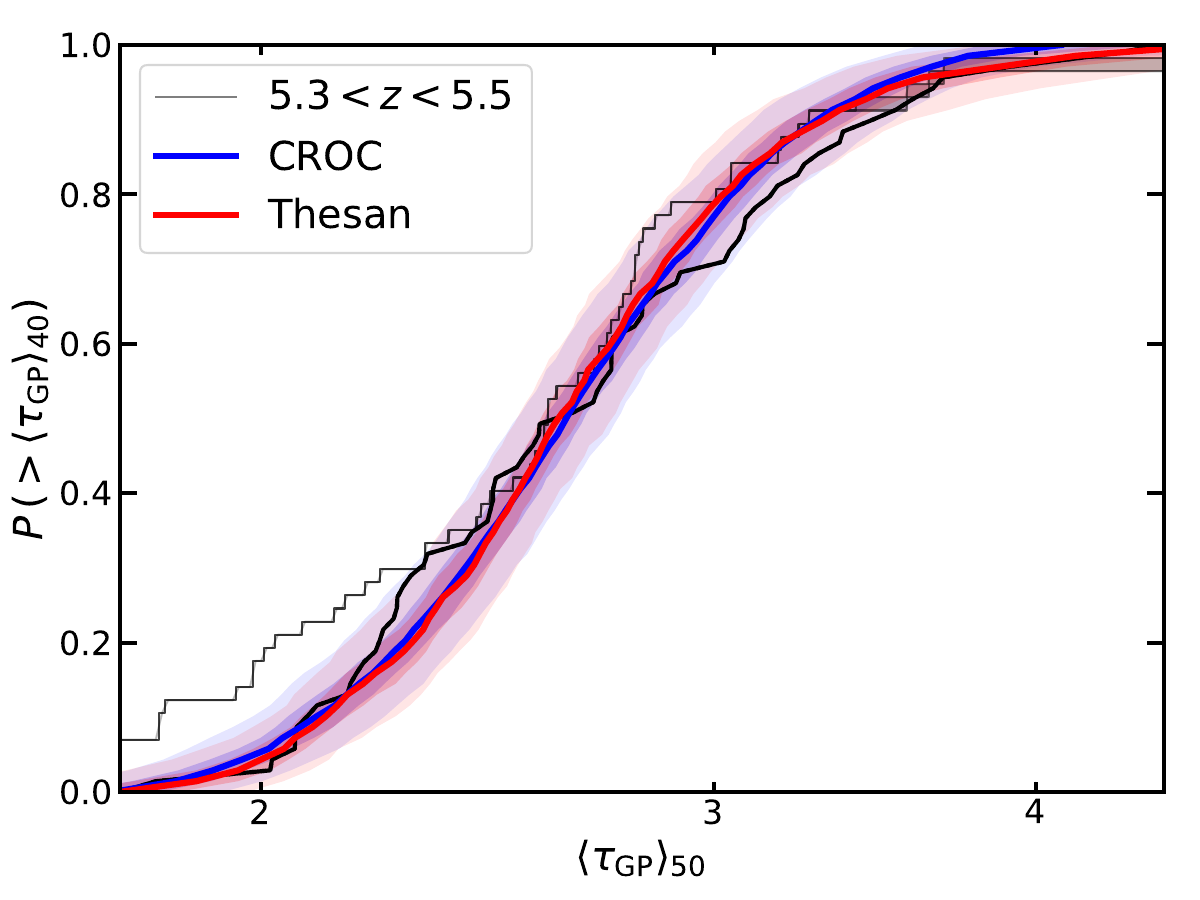}%
\includegraphics[width=0.25\columnwidth]{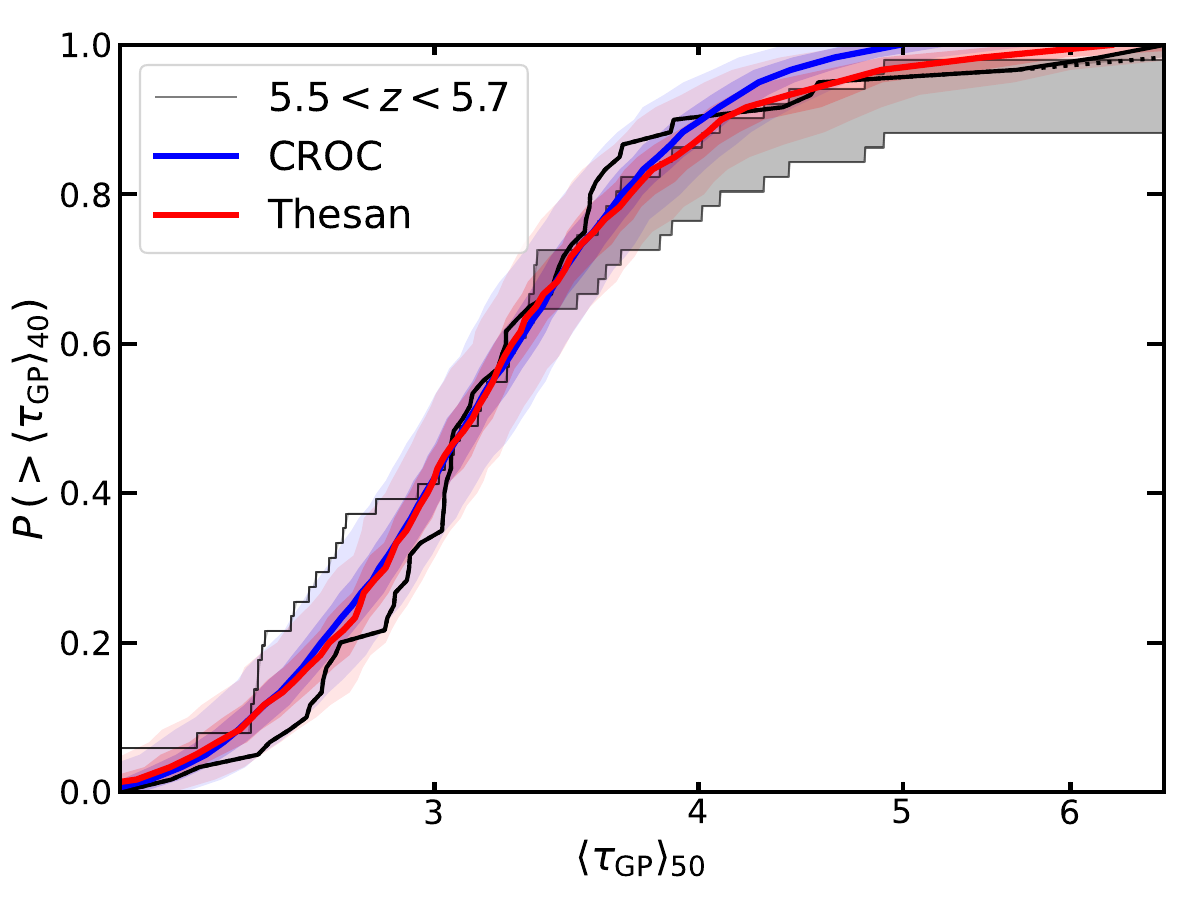}%
\includegraphics[width=0.25\columnwidth]{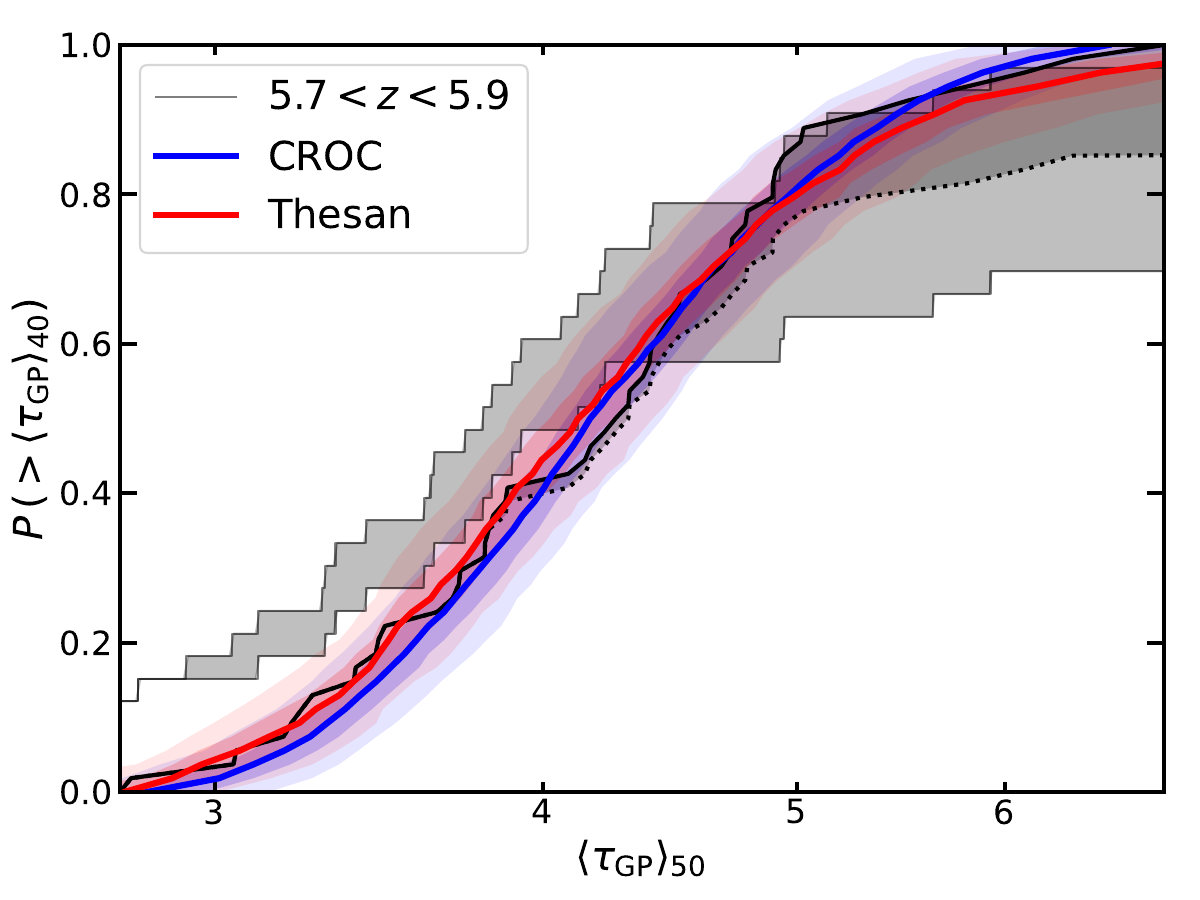}%
\includegraphics[width=0.25\columnwidth]{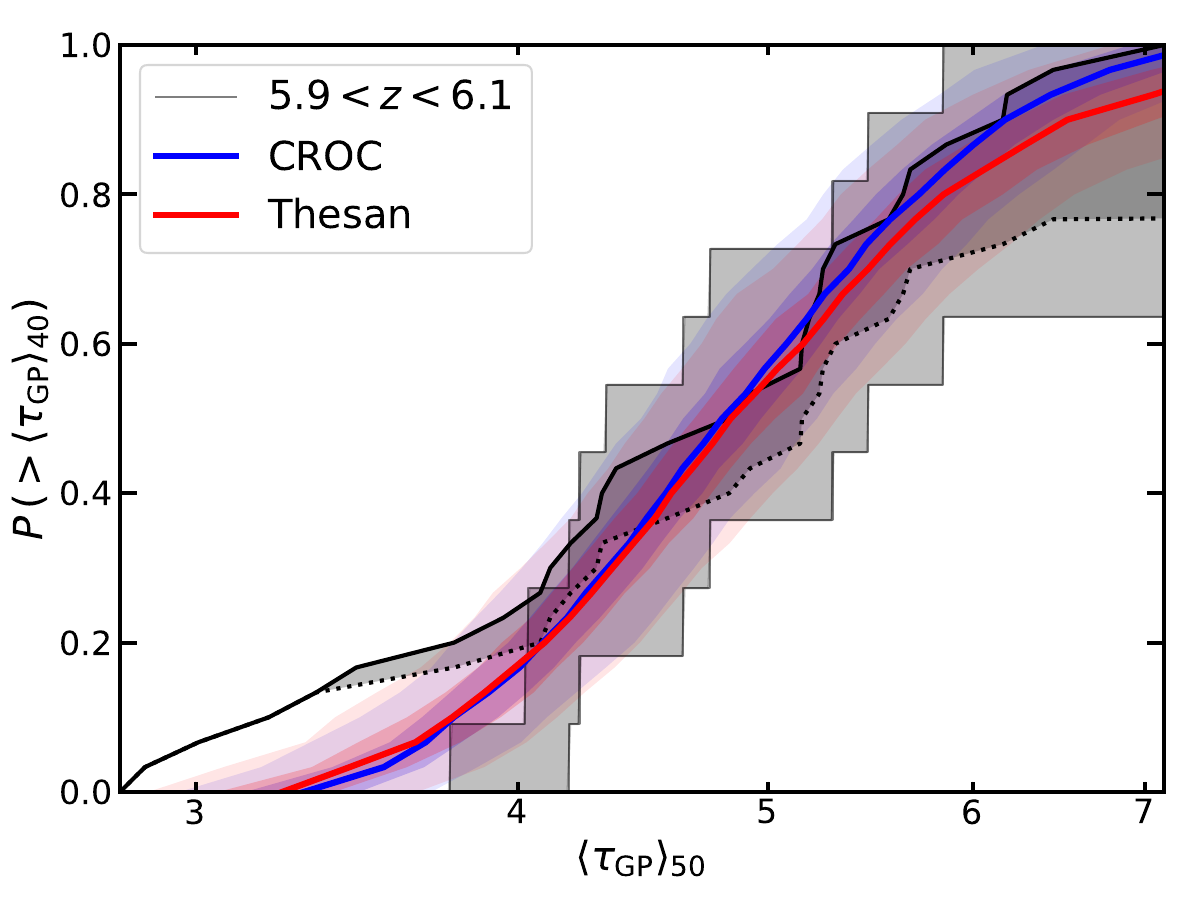}\hfill
\caption{Distributions of effective opacities in CROC and Thesan in 4 redshift bins (red/blue lines) and their sample variance (red/blue semi-translucent bands). Black lines/bands are the observational data from \citet{Bosman2018} and \citet{Bosman2022}. The top row of panels shows the original simulation data. In the bottom row, the simulation data are shown in redshift bins of the same size ($\Delta z=0.2$) but slightly shifted to provide a good match to observations. The values of the bins are listed in Table \ref{tab:ptau}.}
\label{fig:ptau}
\end{figure}

\begin{table}[H]
\caption{Centers of best-matching redshift bins in the simulations and data}
\centering
\begin{tabular}{ccc}
    Observations & CROC & Thesan\\
    \hline
    5.4 & 5.43 & 5.55 \\
    5.6 & 5.60 & 5.73 \\
    5.8 & 5.88 & 5.96 \\
    6.0 & 6.05 & 6.10 \\
    \hline
\end{tabular}
\label{tab:ptau}
\end{table}

Matching between distributions of effective opacities in CROC and Thesan and observations \citep{Bosman2018,Bosman2022} are shown in Figure \ref{fig:ptau} and Table \ref{tab:ptau}.

\clearpage

\section{Iliev Tests for ALTAIR Radiative Transfer Solvers}
\label{app:c}

\begin{figure}[H]
\centering
\includegraphics[width=0.55\columnwidth]{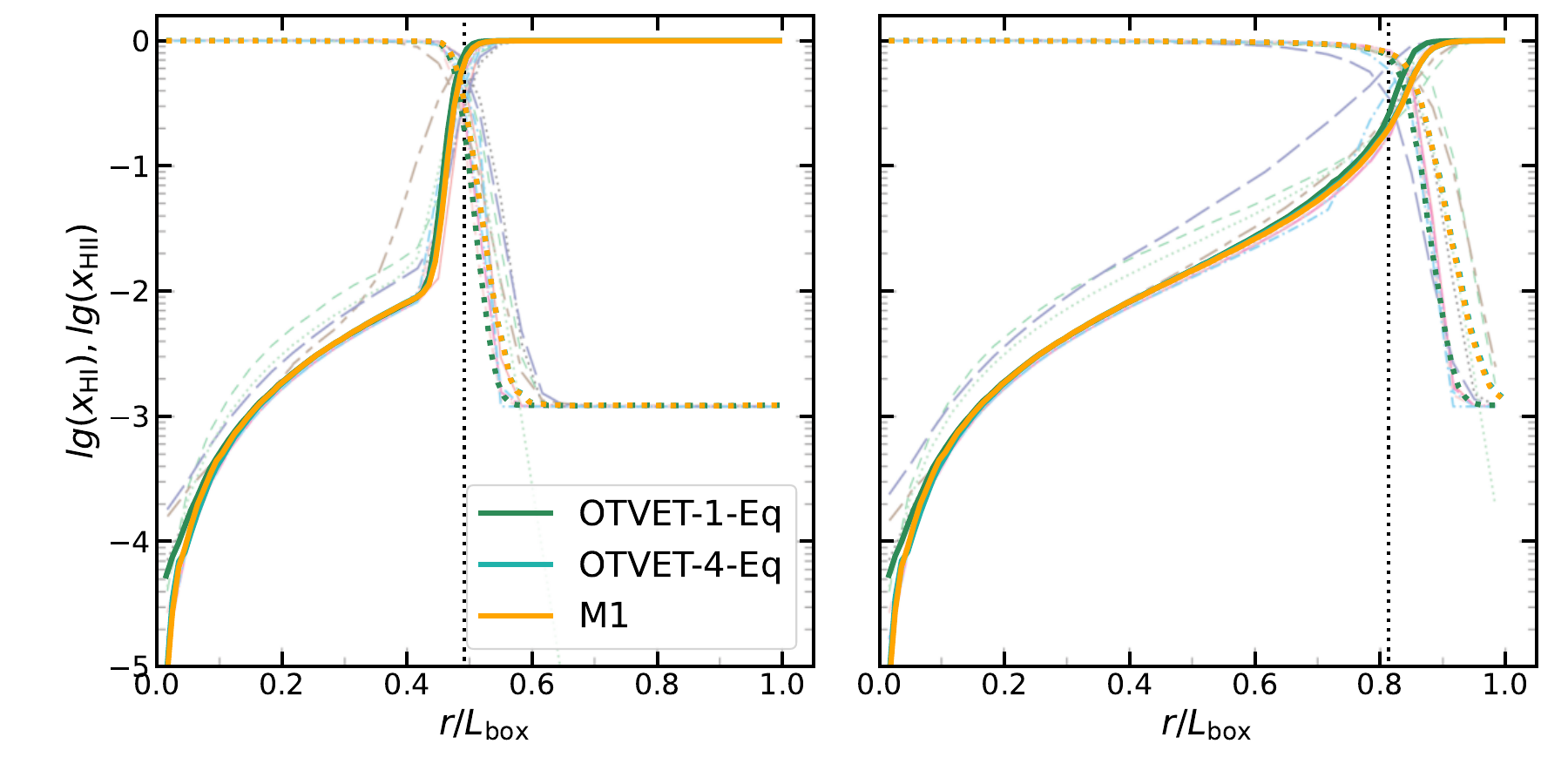}%
\caption{Test 1 from \citet{iliev1} for the three ALTAIR solvers. Line colors match those in Figure \ref{fig:alt4}. The background of this image is the corresponding plot (Figure 8) from \citet{iliev1}, which illustrates the spread among the different RT solvers tested in the original paper.}
\label{fig:iliev1}
\end{figure}

\begin{figure}[H]
\centering
\includegraphics[width=0.7\columnwidth]{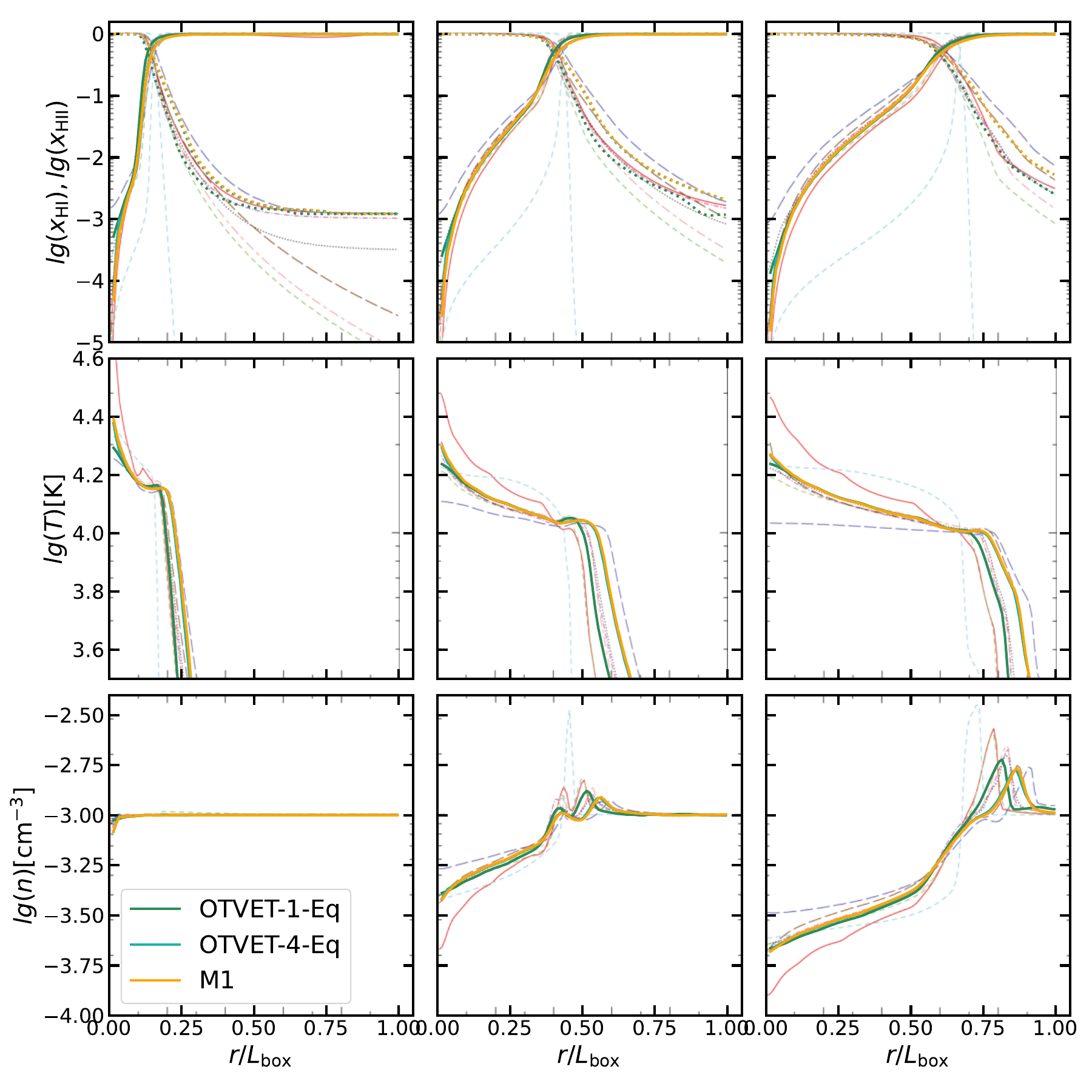}%
\caption{Test 5 from \citet{iliev2} for the three ALTAIR solvers. Line colors match those in Figure \ref{fig:alt4}. The background of this image is the corresponding plots (Figures 13, 15, and 18) from \citet{iliev2}, which illustrate the spread among the different RT solvers tested in the original paper.}
\label{fig:iliev5}
\end{figure}

Figures \ref{fig:iliev1} and \ref{fig:iliev5} show the performance of the three ALTAIR solvers on Iliev tests 1 and 5.

\end{appendix}

\end{document}